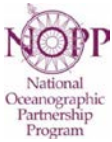 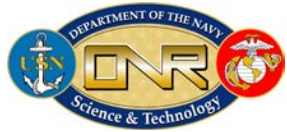 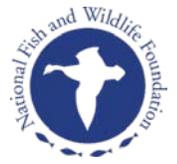

# DCL System Using Deep Learning Approaches for Land-Based or Ship-Based Real-Time Recognition and Localization of Marine Mammals

**Peter J. Dugan**
Bioacoustics Research Program
Cornell Laboratory of Ornithology
Cornell University
159 Sapsucker Woods Road, Ithaca, NY 14850

phone: 607.254.1149     fax: 607.254.2460     email: pjd78@cornell.edu

**Christopher W. Clark**
Bioacoustics Research Program
Cornell Laboratory of Ornithology
Cornell University
159 Sapsucker Woods Road, Ithaca, NY 14850

phone: 607.254.2408     fax: 607.254.2460     email: cwc2@cornell.edu

**Yann André LeCun**
Computer Science and Neural Science
The Courant Institute of Mathematical Sciences
New York University
715 Broadway, New York, NY 10003, USA

phone: 212.998.3283     mobile phone: 732.503.9266     email: yann@cs.nyu.edu

**Sofie M. Van Parijs**
Northeast Fisheries Science Center, NOAA Fisheries
166 Water Street, Woods Hole, MA 02543

phone: 508.495.2119     fax: 508.495.2258     email: sofie.vanparijs@noaa.gov

**LONG-TERM GOALS**

The ONR DCL grant focuses on advancing the state of the art for bioacoustic signal detection and classification through researching new technologies, algorithms and systems. This work engages a unique team of experts from Cornell University (CU), New York University (NYU) and Northeast Fisheries Science Center (NEFSC). Aimed at developing new methods and practices for advancing detection classification (DC), the grant team also maintains a higher level goal: addressing general data mining strategies as applied to large, complex acoustic datasets. The underlying focus of the team's work is to integrate and develop new technologies for hardware and software tools based on high performance computing, and reduce these to practice through outreach into the broader bioacoustic community.

**OBJECTIVES**

The scope of this work includes technical and applied objectives. Technically, the team refined hardware and software tools for high performance processing of sounds. The hardware is referred to as the High Performance Computing, Acoustic Data Accelerator (HPC-ADA) and the software toolset, based on time series acoustic signal Detection cLassification and Machine learning Algorithms, is called DeLMA. Both HPC-ADA and DeLMA had several advances which are discussed in this report. Applied objectives included research into two types of signal catagories. Short frequency modulated sounds comprise the first signal type (TYPE-I). In connection with the international machine learning community, new algorithms specific to right whale (*Eubalaena glacialis*) up calls were developed and applied to large datasets. Algorithms for a second signal type (TYPE-II), consisting of repeating pulse-like signals were investigated. Software prototypes were developed and integrated into distinct streams of reseach; projects include data processing for detecting minke whale (*Balaenoptera acutorostrata* ) and seismic survey modeling. A new approach based on human intuition was developed. The Both TYPE I and TYPE II algorithms were integrated into the HPC-ADA and DeLMA technologies and applied to several research efforts to study complex sound archives spanning large spatial and temporal scales. A new post processing method for detection and classifcation was also investigated. The new method combines human knowledge with an artificial neural network stage, called HK-ANN, and is used to reduce areas of high false positive rates. HK-ANN was successfully tested for a large minke whale dataset, but could easily be used on other signal types. Various publications asscoiated with these objectives were either submitted or published; these are summarized in this report.

**APPROACH**

The work focuses on algorithms, hardware and software for advancing detection, classification and general data-mining capabilities, such as localization, as applied to scalable data environments. All software and hardware are developed using commercial off the shelf components (COTS). The software model was developed using MATLAB and is designed to scale from a laptop (or desktop) application to large, distributed server hardware platforms. Design includes a flexible HPC interface, capable of running a variety of algorithms concurrently, across multiple datasets and sound formats. Hardware system was developed as a powerful distributed server platform for executing big data applications for single, or multi-channel datasets.

*Dugan, Clark, LeCun and Parijs*

**WORK COMPLETED**

The research achieved several significant outcomes. First, the new software and hardware framework for procesing sounds using high performance computing (HPC) technologies continue to be improved from earlier phases. The DeLMA HPC software, uses a generalized time-series algorithm model specifically optimized for processing sounds. The model is flexible, allowing integration of customized data-mining algorithms, and scalable, operating on platforms ranging from laptops to distributed server systems. Research shows that the processing model can support a range of inputs consisting of different data sources and sensor types, addressing hybrid-systems and complex data needs. Configurable load balancing and remote-distributed processing offers the capability of supporting ocean-scale computing while utilizing the power of HPC. DeLMA was recently enhanced to accommodate several new algorithms and projects. Major enhancements include a graphical user interface and object oriented structure, allowing for easier integration on large server-based computing platforms as well as client based environments. Current work is underway and investigates the practical use for scaling to client-server environments, whereby laptops (or desktop) platforms can accommodate seamless computing operations, gaining access to a large pool of distributed workers for processing sound archives. Figure 1(a) shows the newly created HPC graphical user interface, capable of running a variety of algorithms concurrently, across multiple datasets and sound formats.

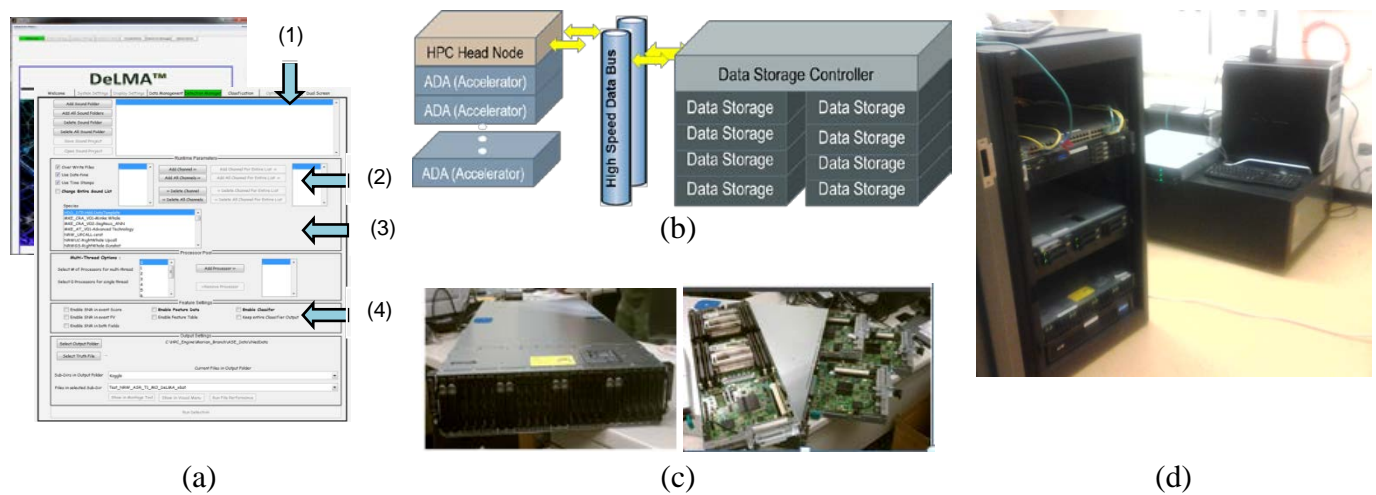

Figure 1. (a) HPC DeLMA software: Users can (1) add large sound archives, (2) configure the input sensor, (3) configure data mining algorithms and (4) load balance through processor selection. (b) Dell- C6220 64 core server. (c) HPC-ADA system.

Figure **1**(b) shows recent consolidation of the hardware referred to as the HPC Acoustic Data Accelerator (HPC-ADA). Hardware consolidation has enabled a powerful connection between processing accelerators and a scalable network attached storage device (s-NAS). Design uses a scalable approach for both processing resources and data storage. Processors scale using the HPC Server hardware (Figure 1[c]), whereby each processor module, or accelerator, can be added to the design for increased power. For example, Figure 1(c) shows four accelerators, each having 16 processors mounted in a single Dell C6220 server. Adding another C6220 module would gain

*Dugan, Clark, LeCun and Parijs*

additional 64 processors, totaling 128 usable worker cores. Data scalability is realized by the recent addition of a 48 TB NAS, which is connected four ways through the high-speed network switch. Additional 48 TB was recently added to the current design, reaching the 96 TB level. Each storage unit has a dedicated network attachment, providing additional performance.

The complete system, Figure 1(d), shows integrated hardware and software. The software (DeLMA) has a special interface providing total distribution of algorithms and sounds, making use of the scalable dedicated network architecture. Software is also functionally independent of changes in the hardware structure, meaning that DeLMA does not need to be modified when NAS or accelerator changes are made. Instead, the software automatically senses additional hardware, allowing the user to selectively change computing resources from the window interface shown in Figure 1(a). Work was also invested to provide batch-processing capability. Batch processing is critical for big data applications, making use of pre-configured jobs to execute unique algorithms across projects with dissimilar sensor and sound structures. For example, newly designed right whale algorithms can execute on projects measured across multiple years, with each project configured to a separate channel and sample rate.

The HPC-ADA server hosts the DeLMA software application with several data-mining algorithms to support research in which Cornell collaborated with institutions and businesses to promote various projects. Project requirements consisted of having two different types, or categories, of sounds that needed detection-classification algorithms: TYPE-I, described by short duration and highly variable, frequency-modulated (FM) sounds, and TYPE-II, longer repeating pulse trains (PT) with semi-regular behavior. All TYPE-I work focused on up-call sounds generated from the North Atlantic Right Whale (NARW). Signals were extracted from regions on the eastern coast of the United States. Early phases of TYPE-I research collaborated with NYU to develop a working prototype of a biologically motivated algorithm based on the convolutional neural network (CNN). Research was further explored within ICML through Kaggle.com and the bioacoustics community. Cornell developed two new TYPE-I algorithms, one based on connected region processing (CRA) and the second on a histogram of oriented gradients (HOG). Both CRA and HOG methods were fitted to the DeLMA software for scalable processing. The team competed in two international competitions and hosted a third competition at the Bioacoustics Workshop at the International Conference on Machine Learning (ICML-2013). TYPE-II algorithm work was supported and applied to several existing and new projects, supporting detection for several signal types including a variety of human-made (seismic) and mysticete sounds. Basic components of the pulse train algorithm contain three main phases: aggregation, segmentation and registration, referred to as ASR pulse train or ASR-PT. ASR technology was supported in various projects including seismic survey and several minke whale studies.

A new "post processing classifier" was investigated that can be modified for any algorithm with an exposable feature vector called the human knowledge artificial neural network, (HK-ANN). Since the user could rapidly run large datasets using the HPC-ADA, the goal was to investigate whether a human could see obvious errors over wide temporal and spatial scales and "assist the computer" to correct for a better recognition performance. This concept is based on utilizing the many hours and experience that a single user has, looking at many signal outcomes and *a priori* knowledge at known behavior patterns, such as seasonality and diel activity. The study was based on a simple idea that an expert scientist can incorporate human judgment to a series of measures, or features, thereby forming a more accurate result. The basic process is shown in Figure 2. ASR-PT algorithm is used to extract candidate detections. Using each detection form the ASR-PT block in Figure 2, corresponding features along with an image of the event are created for visual observation. The user inspects a series of

*Dugan, Clark, LeCun and Parijs*

images. For each image, the expert assesses the quality of the signal, and factors in knowledge about the diel pattern and seasonal likelihood. After assessment, the user assigns a score, which serves as the "intuition" metric. Basic features such as pulse width, inter-pulse-interval, et cetera, see [9], for each signal are also exported. Feature vector and scores are combined using a neural net training procedure, forming a post classifier algorithm, or HK-ANN stage. The HK-ANN is run on the entire dataset, factoring in expert scores and general feature data for improved error rejection.

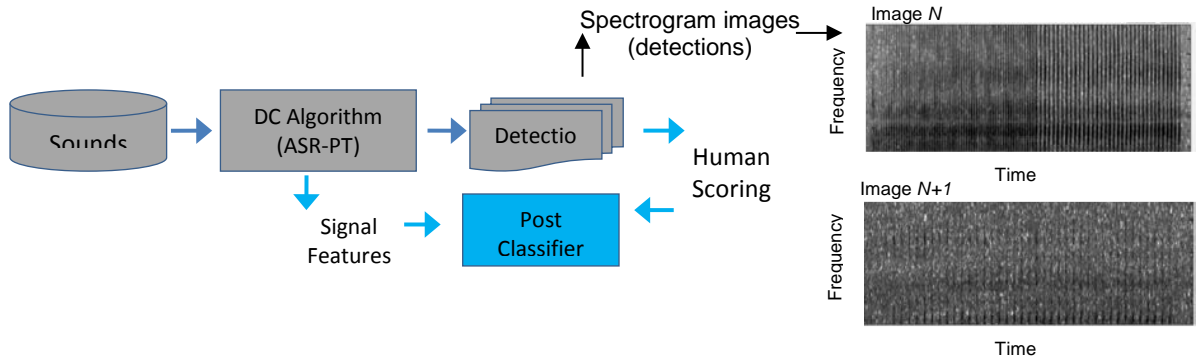

Figure 2. Human knowledge, artificial neural network process (HK-ANN). Operators assign scores to previously detected signal events. Two different pulse train events shown above for example. Signal features are carried forward from the original detection algorithm and combined with human scores to create a "post classifier". The concept is to use a human to assist the computer in detecting proper signal patterns.

## RESULTS

*HPC-ADA, DeLMA Hardware, Software*

The HPC-ADA prototype system was constructed utilizing a DELL Cloud Server C6220, Figure 1. Collectively the platform contains 64 physical cores of Intel Xeon E-2670 @ 2.6 GHz processor, 192 GB of local memory, 2 TB of local disk storage used for local cache, 32 TB of gigabit connected network attached storage for sound archives and data products. The recent addition of a 48 TB expansion slot successfully hosts additional data. From 2011-2013, HPC technology developed through this work has effectively supported 19 large projects at the Cornell University Bioacoustics Research Program and has executed over 3.6 million channel hours of sounds [1]. On all accounts, HPC technology has provided highly desirable data products for all the applied projects. Usability and funding was one drawback. With complex, expensive systems, more experienced data scientists were required to use the HPC technology requiring an added cost layer for processing data. These challenges could be addressed by offereing a wider technical community to use the tools, taking advantage of a larger collaboration of projects. It is expected that increasing use by the community will ultimately advance tools in the direction of non-data scientists.

*HPC Seismic Airgun Study*

The team's work also generated a paper on using HPC technology for applying an advanced algorithm based on large data aggregation segmentation and registration (ASR). The seismic airgun study measured over 160,000 pulses during a 42 day seismic survey. See Figure 3 (measurements taken using the HPC-ADA machine). The work coupled the ASR algorithm with the HPC-ADA system, achieving rapid performance, and processing the complete siesmic archive in 3.75 days, versus 45 days required by a traditional serialized methods. This work showed that a hybrid set of data could be combined with HPC processing to develop a large collection of acoustic measures, including estimates of source levels, receive levels and various sound pressure metrics. A peer review manuscript is currently under review [2].

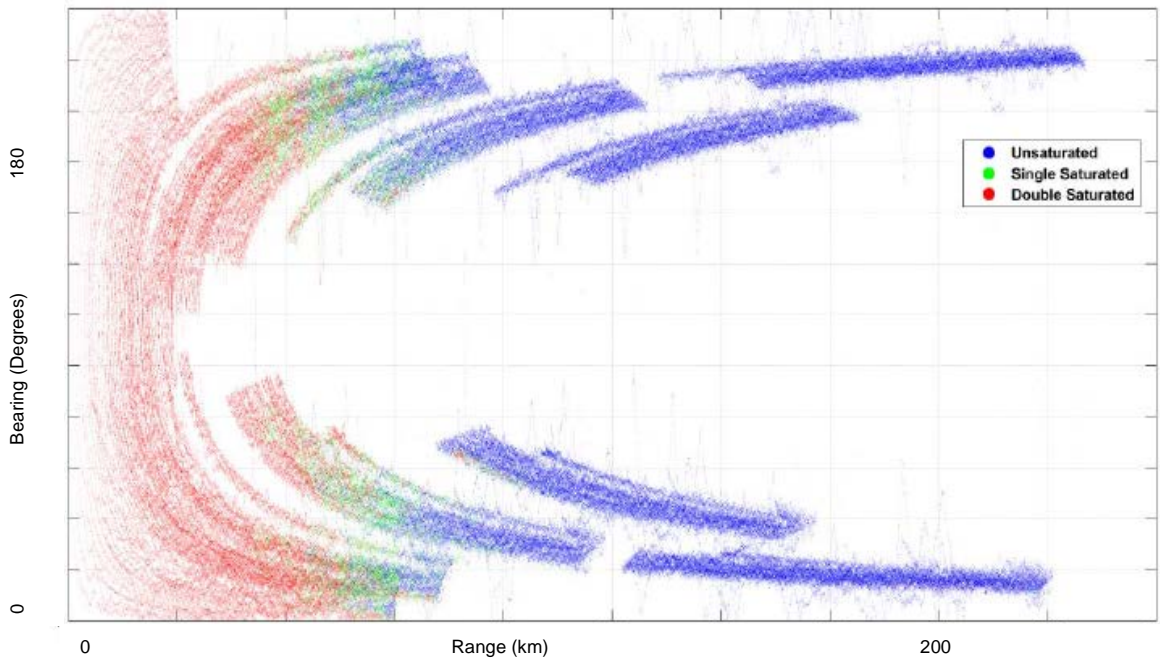

Figure 3. HPC-ADA managed over 160,000 seismic pulses during 42 day study in Baffin Bay, Greenland. Data measured with respect to the air-gun vessel, color showing saturation modes with respect to bearing and range.

*TYPE I, Automatic Recognition Study: Right Whale*

Another accomplishment is the automatic recognition project, performed on the HPC tools using algorithms from 2013 Kaggle competition. The team compared three algorithms designed to detect and classify North Atlantic right whale context calls (upcalls). Automatic recognition was compared against earlier work [3], which had been verified through human observation. The work demonstrated that base line algorithms provied a high degree of error (false positives) and did not show any visual evidence of seasonal patterns. The two algorithms utilized by the team from the Kaggle 2013 competition include connected region analysis (CRA) [4] and histogram of oriented gradients (HOG). Both methods demonstrated seasonal patterns that agreed with earlier studies, work published in [5].

*Dugan, Clark, LeCun and Parijs*

*TYPE-II, Pulse Train Algorithms using ASR and Deep Scattering methods*

Other advances were logged in the ASR multi-stage seismic algorithm, developed for Minke Whale and succesfully used on various large projects, [1, 6]. The multi-stage algorithm incorporated in the HPC Work was summarized in a manuscript for publication [7]. Cornell and New York University collaborated on a new algorithm for detecting repeating energy signatures. The goal of this work was to combine concepts from deep learning with pulse-like signals, TYPE II. The initial work and algorithm was completed along with small sample test runs. Further development to integrate this promising work with the HPC technology is ongoing.

*Human Knowledge Artificial Neural Network (HK-ANN) Algorithm*

The study used a large, multi-seasonal minke whale result from research efforts in [8]. Advances were made in an experimental study performed to investigate the practical applications for having users interface with large data systems (such as the HPC-ADA) for reducing detection errors by using operator assisted, "Post Classifier". Figure 4 shows detection performance comparing three years of diel plots for the PT-Algorithm, Human Expert and HK-ANN.

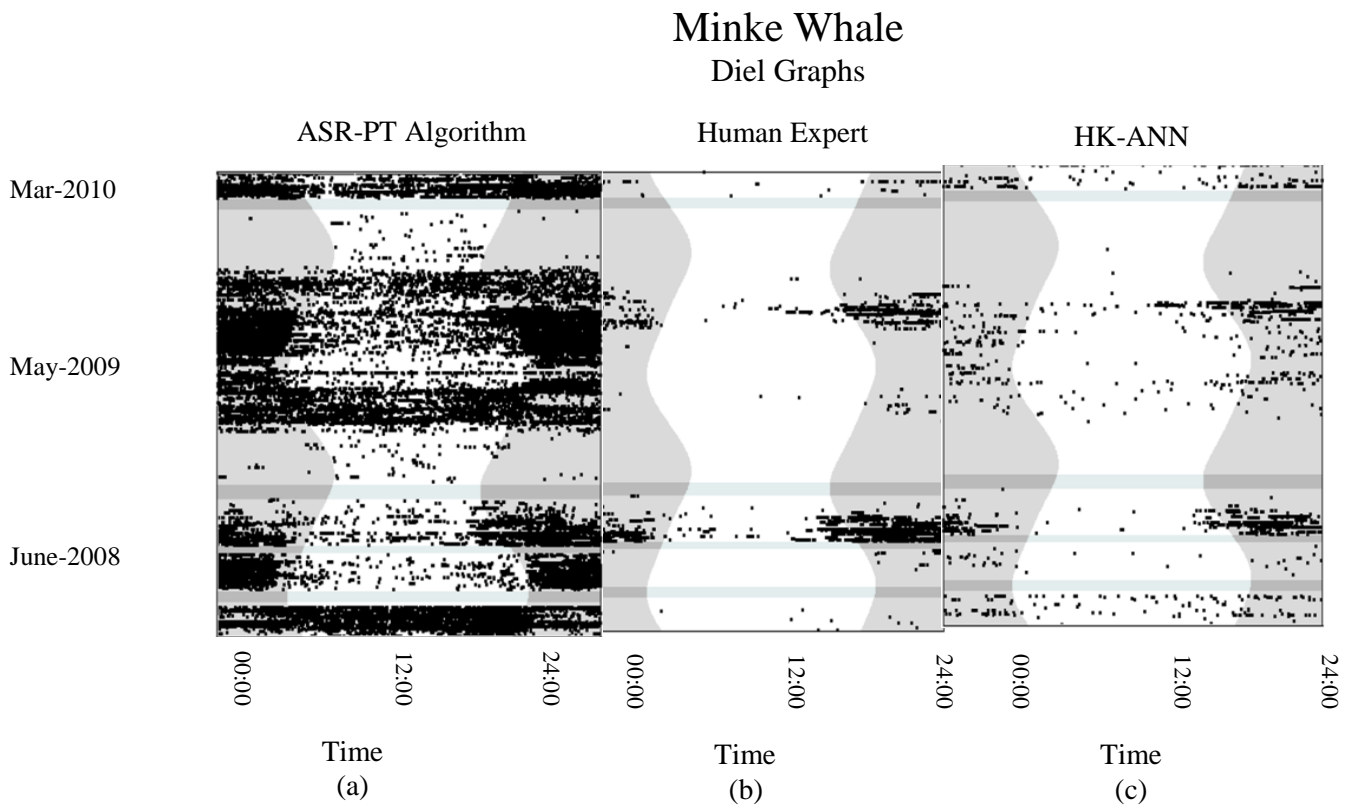

Figure 4. Diel results showing (a) ASR-PT algorithm, (b) Human Expert hand truth and (c) results of the AK-ANN. Visual inspection shows a high degree of similarity between Human Expert and the HK-ANN; where the ASR-PT has a high detection concentration, or errors, indicated by the dense number of events.

In total, 41,560 Minke events Figure 4(a) were generated by original ASR algorithm [7]. To minimize fatigue, only 2,625 events were scored by the expert user, identifying errors using a scale ranging from 1 through 5. HK-ANN was compared against three different classifiers consisting of Bayesian Network, Grafted Decision Tree and Classification Regression Tree. Results shown in Figure 5 indicate a significant improvement in overall performance, especially at low False Positive Rates (FPR), by using the proposed HK-ANN method. For example, at a FPR of 6% we have an improvement in True Posiitve Rate (TPR) of approximately 20%. More details for the work can be found in [9].

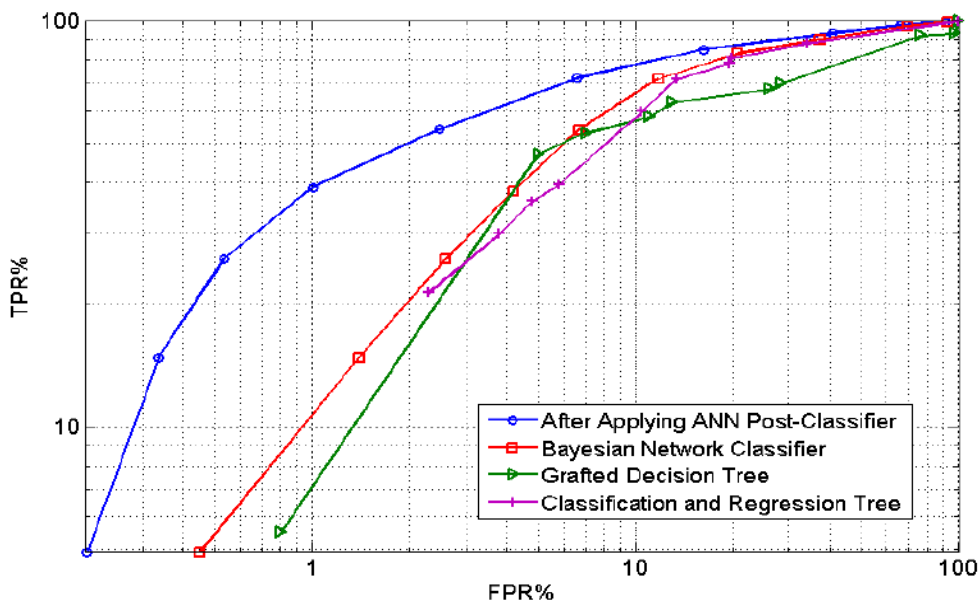

Figure 5. Receiver operator curve comparing HK-ANN post processing classifier to three inline classifiers consisting of Bayesian Network, Grafted Decision Tree and Classification Regression Tree.

*Applied Biological Studies*

ASR-Minke pulse train algorithm was further made available for running sound archives at CU; with Northeast Fisheries Science Center (NEFSC) performing several indepth studies for the minke whale. Using passive acoustic datasets spanning large temporal and spatial scales, studies centered around multi-year seasonal and diel vocalization [8], seasonal migrations [10] and individual calling behaviour and movements [11].

**IMPACT/APPLICATIONS**

HPC-ADA machine and DeLMA software are models which have been successfully used with algorithms beyond detection-classification, including noise analysis and acoustic modeling. Client-server architectures have also been explored through application of the MATLAB Distributed Computing Server (MDCS). Further applications and research for this work are ideal for leveraging systems requiring large, complex data-mining operations. Further investment for this work should be done at a larger scale within the bioacoustic community, offering access to a wider breadth and diversity of analysis projects.

*Dugan, Clark, LeCun and Parijs*

**PUBLICATIONS**

P.J. Dugan, J.A. Zollweg, H. Glotin, M.C. Popescu, D. Risch, Y.A. LeCun and C.W. Clark (2014), "High Performance Computer Acoustic Data Accelerator (HPC-ADA): A New System for Exploring Marine Mammal Acoustics for Big Data Applications," ICML 2014, Workshop on Machine Learning for Bioacoustics, Beijing, China, *in press*.

P.J. Dugan, M. Guerra, D.W. Ponirakis, J.A. Zollweg, M.C. Popescu and C.W. Clark (2014), "Seismic Air-gun Survey, Fine Resolution, Passive Acoustic Data Measurements Using High Performance Computing: Case Study Baffin Bay," *IEEE Journal of Oceanic Engineering*, November 13, 2014, *in review*.

M.C. Popescu, P.J. Dugan, C.W. Clark and A.N. Rice (2014), "An Automatic Approach to Marine Mammal Pulse Type Acoustic Signature Recognition," *J. Acoustical Society of America*, November, *in review*.

M. Pourhomayoun, P.J. Dugan, M.C. Popescu and C.W. Clark (2013), "Bioacoustic Signal Classification Based on Continuous Region Processing, Grid Masking and Artificial Neural Network, *ICML 2013 Workshop on Machine Learning for Bioacoustics* , arXiv preprint arXiv:1305.3635.

D. Risch, C.W. Clark, P.J. Dugan, M.C. Popescu, U. Siebert and S. Van Parijs (2013), "Minke whale acoustic behavior and multi-year seasonal and diel vocalization patterns in Massachusetts Bay," USA, Mar Ecol. Prog. Ser. 489:279-295.

D. Risch, M. Castellote, C. Clark, G. Davis, P. Dugan, L. Hodge, A. Kumar, K. Lucke, D. Mellinger, S. Nieukirk, M. Popescu, C. Ramp, A. Read, A. Rice, M. Silva, U. Siebert, K. Stafford and S. Van Parijs (2014), "Seasonal migrations of North Atlantic minke whales: Novel insights from large-scale passive acoustic monitoring networks," *Movement Ecology*.

D. Risch, U. Siebert and S. Van Parijs (2014), "Individual calling behavior and movements of North Atlantic minke whales (*Balaenoptera acutorostrata*)," vol. 151, no. 9, pp. 1335-1360.